# Uncovering hidden spin polarization of energy bands in antiferromagnets

Lin-Ding Yuan[1], Xiuwen Zhang[1], Carlos Mera[2], Alex Zunger[1]

[1]Renewable and Sustainable Energy Institute, University of Colorado, Boulder, Colorado 80309, USA

[2]Center for Natural and Human Sciences, Federal University of ABC, Santo Andre, São Paulo, Brazil

## Abstract

Many textbook physical effects in crystals are enabled by some specific symmetries. In contrast to such 'apparent effects', 'hidden effect X' refers to the general condition where the nominal global system symmetry would disallow the effect X, whereas the symmetry of local sectors within the crystal would enable effect X. Known examples include the hidden Rashba and/or hidden Dresselhaus spin polarization that require spin orbit coupling, but (unlike the apparent Rashba and Dresselhaus counterparts) can exist even in inversion-symmetric non-magnetic crystals. Here we point out that the spin splitting effect that does not require spin-orbit coupling (SOC) can have a hidden spin polarization counterpart in antiferromagnets. We show that such hidden, SOC-independent effects reflect intrinsic properties of the perfect crystal rather than an effect due to imperfections, opening the possibility for experimental realization, and offering a potential way to switch antiferromagnetic ordering.

**Introduction**

*Apparent effects enabled by the symmetry of the global system:* Many traditional textbook physical effects in crystals are enabled by some specific symmetries, encoded in the crystal space group. Such are the symmetry conditions for the apparent electric polarization which defines various order parameters such as in ferroelectricity [1], circular dichroism [2], and pyroelectricity [3]. Another example of effects enabled by the recognized global system symmetry is remove of spin degenerate energy bands (spin splitting) due to spin-orbit coupling (SOC) induced split energy bands in non-magnetic crystals having broken inversion symmetry (such as the Rashba (R-1)[4] and Dresselhaus (D-1)[5] effects). When an effect is observed despite the needed enabling symmetry being absent, it is traditional to assume that the system contains some symmetry-altering imperfections.

*Hidden Effect X:* Here we discuss "Hidden effect X" that occurs in systems that do not support by the nominal enabling symmetry, yet effect X exists locally and reflects intrinsic properties of the perfect crystal rather than imperfections that would disappear when the crystal becomes perfect. The understanding of such hidden intrinsic effects is important as it can demystify peculiar observations of phenomena that are unexpected to exist based on the global symmetry of the system.

*Examples of "Hidden effect X" that is SOC-induced:* These include (i) X= "anisotropic optical circular polarized luminescence", expected only in non-centrosymmetric crystals but observed [6] also in centrosymmetric transition-metal dichalcogenides, originally dismissed as being due to some extrinsic sample imperfection [7-9] but later on it was shown to be an intrinsic property pertaining to the individual layer [10]. (ii) Rashba or Dresslhause spin polarization, expected exclusively to occur in non-centrosymmetric crystal, but predicted [11,12] and observed [13-23] in centrosymmetric nonmagnetic crystal (denoted R-2 and D-2, respectively). A similar form of Hidden effect X includes (iii) X= "spin polarization" induced by SOC was studied for antiferromagnetic systems. The effect is again expected only in non-centrosymmetric compounds but shown in "centrosymmetric" antiferromagnets CuMnAs and $Mn_2Au$ [24,25] where the combined symmetry of inversion and time reversal disallows splitting. Prominently, the hidden spin polarization in these compounds facilitates the electrical reversal of their antiferromagnetic ordering [24,25].

*Apparent and hidden effect X that is SOC independent:* Unlike the above noted SOC-induced apparent and hidden effect, here we discuss a different form of hidden effect whose corresponding apparent effect is independent of SOC and exists in antiferromagnetic materials where spin-up and spin-down bands are paired. Such "SOC-independent hidden spin polarization" are illustrated in Fig. 1(a). We delineated, based on symmetry, six types of such hidden effects. This complements to existing picture of R-2 and D-2 in non-magnetic solids. We identified many material candidates and provided Density functional theory results for several example compounds showing that the SOC-independent hidden effects are intrinsic to the bulk. This opens the possibility for experimental realization, electric field control of the hidden effect, and potential new ways to switch antiferromagnetic ordering.

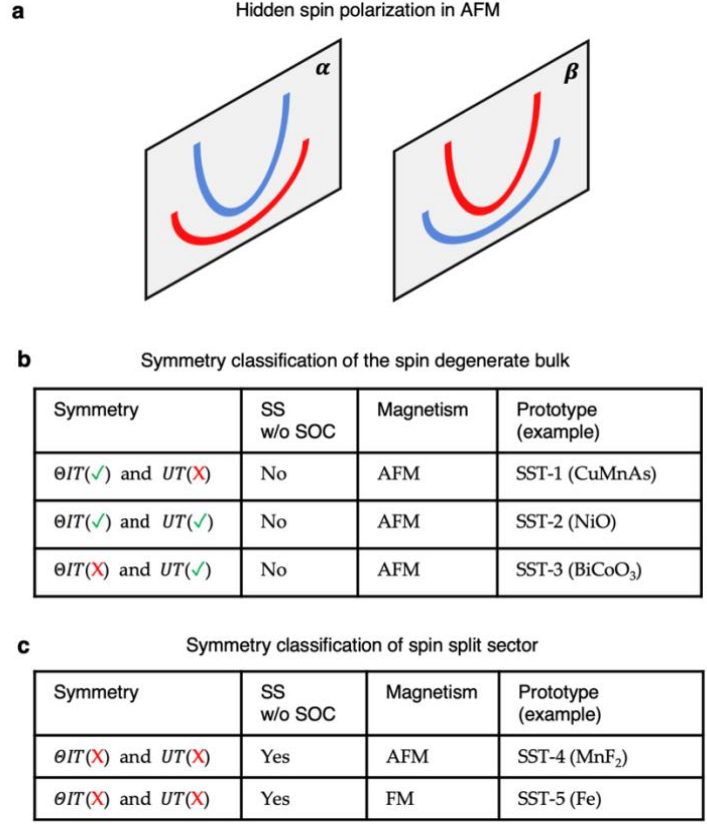

**Figure 1: Hidden spin polarization in collinear antiferromagnets without SOC. a** Schematic illustration of SOC-independent hidden spin polarization; **b** Symmetry conditions for spin degenerate bulk; **c** Symmetry conditions for spin split sector. In panel (a), sectors are represented by color-shaded planes, the red and blue lines in the plane represent the spin-up and spin-down bands. In panel (b) and (c) checkmark and cross in parentheses are used to indicate the presence or absence of the symmetry it follows for all panels.

## Results

### Enabling symmetry conditions for SOC-independent apparent spin polarization in antiferromagnets

Symmetry is essential to understand the energy bands' degeneracy of a material. The symmetry conditions for apparent spin splitting or spin polarization was pointed out recently in Ref. 26. Such symmetry conditions disentangle the SOC-independent splitting from the SOC-induced splitting by considering the symmetry at the zero SOC limit [27-29], where spin and space are fully decoupled. This involved utilizing first a few individual symmetry operations: $U$ being a spin rotation of the SU(2) group acting on the spin 1/2 space that reverses the spin; $T$ being spatial translation; $\theta$ being time reversal, and $I$ being the spatial inversion. These individual operations are then used for constructing two symmetry products: a SOC-free magnetic symmetry $\theta IT$, and a spin symmetry $UT$ (where the former product can be simplified to $\theta I$ by proper choice of inversion center). SOC-independent spin splitting [26,30] would occur only when both symmetry products are simultaneously violated. "Centrosymmetric antiferromagnets" [31] with $\theta IT$ symmetry will not show such spin splitting.

Given the symmetry conditions, it is thus possible to classify all different spin splitting prototypes [26,30] for magnetic materials. There are three prototypes with no apparent spin splitting effect: (1) AFM compounds that violate $UT$ but preserve $\theta IT$ symmetry referred to as spin splitting prototype 1, referred to as SST-1 antiferromagnets; (2) AFM compounds that preserve both $UT$ and $\theta IT$ symmetry referred to as SST-2 antiferromagnets; (3) AFM compounds that preserve $UT$ but violate $\theta IT$ symmetry referred to as SST-3 antiferromagnets.

In addition, there are two prototypes with apparent spin splitting effects: (4) AFM compounds that violate both $UT$ and $\theta IT$ symmetry referred to as SST-4 antiferromagnets; (5) Ferromagnetic (FM) compounds that violate both $UT$ and $\theta IT$ symmetry referred to as SST-5 ferromagnets. The classification is summarized in Fig. 1b,c. In the following, we will use these notations to describe the symmetry conditions for the hidden spin polarization effect in antiferromagnets.

## Enabling symmetry conditions for hidden SOC-independent spin polarization in antiferromagnets

"Hidden spin polarization" is expected in collinear antiferromagnets when the bulk has zero net spin polarization, but its constituent sectors allow locally a spin splitting and spin polarization effect. Consider the combination of two possible prototypes constituting sector that gives hidden spin polarization locally but lead to three possible prototypes of the bulk symmetry (preserving either $\theta IT$ or $UT$ or both) that disallows apparent spin polarization, one can then classify six hidden spin polarization cases. Following the previous classification of spin splitting prototypes for apparent spin degeneracy and apparent spin splitting [26,30], collinear antiferromagnetic materials with "hidden spin polarization" are those antiferromagnets whose bulk prototype being SST-I (I = 1, 2, 3) and constitute sector prototype being SST-J (J = 4, 5). Detailed discussions of the symmetry conditions for hidden spin polarization in collinear AFM are given in Supplementary Information Section A.

Figure 2 summarizes the six possible types of hidden spin polarization without SOC in antiferromagnets that are spin degenerate but contain spin split sectors (represented by color-shaped plane). Fig. 2a-c illustrates the three cases where the spin degenerate antiferromagnets of SST-I (I = 1,2,3) can be decomposed into alternating ferromagnetic local sectors that locally violate both $UT$ and $\theta IT$, thus allows spin splitting without SOC. FM materials that satisfy the conditions of violating both $UT$ and $\theta IT$ (always true) are denoted as SST-5 in Fig. 1. The three magnetic-induced hidden spin polarization cases can then be denoted as (a) bulk SST-1 sector SST-5; (b) bulk SST-2 sector SST-5, and (c) bulk SST-3 sector SST-5. Fig. 2d-f illustrates the three cases where the spin degenerate AFM of SST-I (I = 1, 2, 3) can be decomposed into alternating antiferromagnetic local sectors that locally violate both $UT$ and $\theta IT$, thus allows spin splitting without SOC. AFM materials that satisfy the condition are denoted as SST-4 in Fig. 1. The three AFM-induced hidden spin polarization cases can then be denoted as (d) bulk SST-1 sector SST-4; I bulk SST-2 sector SST-4, and (f) bulk SST-3 sector SST-4. We note that there are multiple ways to decompose the bulk system into sectors, e.g. the bulk SST-I (I=1,2,3) might also be decomposed into sector SST-I (I=1,2,3) (or equivalently SST-I (I=1,2,3) sectors can be used to build the bulk SST-I (I=1,2,3) materials), where the local spin polarization of each individual sector is still zero, therefore, are not the focus of this work.

**Figure 2: Six types of SOC-independent magnetic hidden spin polarization in collinear antiferromagnets.** These antiferromagnets have global symmetry that disallows spin splitting without SOC, but have lower local sector symmetry that allows spin splitting without SOC. Cases **a,b,c** is where hidden spin polarization arise from local ferromagnetic sectors and cases **d,e,f** is where the hidden spin polarization arise from local antiferromagnetic sectors. Shaded planes are used to indicate the individual sectors that have neither $\theta IT$ nor $UT$ symmetry and allow spin splitting in the absence of SOC; Parallel and antiparallel arrows of red and blue within the sector plane are used to indicate the ferromagnetic and antiferromagnetic ordering of the sector. Sector symmetry is indicated on top of each plane, and bulk symmetry is indicated by the arrow connecting the two sectors.

## Compounds that have SOC-independent hidden spin polarization

We now turn to discuss how the enabling symmetries are applied to individual sectors to give magnetic hidden spin polarization effects in real antiferromagnetic materials.

As a first step, we will try to find real materials that falls into the six categories we defined. This can be done straightforwardly by applying the symmetry conditions to filter out candidate materials in existing antiferromagnetic databases. We conducted such filtering for MAGNDATA database[32] and identified a few antiferromagnetic materials of potential candidates for magnetic hidden spin polarization. The identified candidates are: $Ca_2MnO_4$ [33], $CoSe_2O_5$ [34] and $Fe_2TeO_6$ [35], $K_2CoP_2O_7$ [36] and $LiFePO_4$ [37] whose bulk prototype is SST-1 with sector prototype of SST-4; $Sr_2IrO_4$ [38] whose bulk prototype is SST-2( meaning trivial) with sector prototype of SST-4; $SrCo_2V_2O_8$ [39] whose bulk prototype is SST-3 ( meaning trivial) with sector prototype of SST-4; CuMnAs [40] and $Mn_2Au$ [41] whole bulk prototype is SST-1 with sector prototype of SST-5; $FeCl_2$ and $CoCl_2$ [42] whose bulk prototype is SST-2 with sector prototype of SST-5; ErAuGe [43] whose bulk prototype is SST-3 with sector prototype of SST-5. These materials form the platform for the exploration of the magnetic hidden spin polarization effects.

The opposite design philosophy (the bottom-to-top approach) is to construct layered bulk antiferromagnets with the hidden effect based on two-dimensional (2D) compounds that belong to SST-4 and SST-5 prototypes. By searching through the database of predicted naturally exfoliate 3D Van der Waals materials [44], we find a list of 37 ferromagnetic 2D materials and 6 antiferromagnetic 2D monolayers that can be used as such building blocks (see Supplementary Information Tables SI and SII for the list). Other predicted and synthesized layered 2D materials are either hypothetical or contradictory to enabling symmetry conditions for AFM spin splitting. Van der Waals compounds with spin splitting not only allow the potential practical controllability through external electric fields but also a platform to explore the coexistence of Van der Waals materials properties and AFM-induced spin splitting.

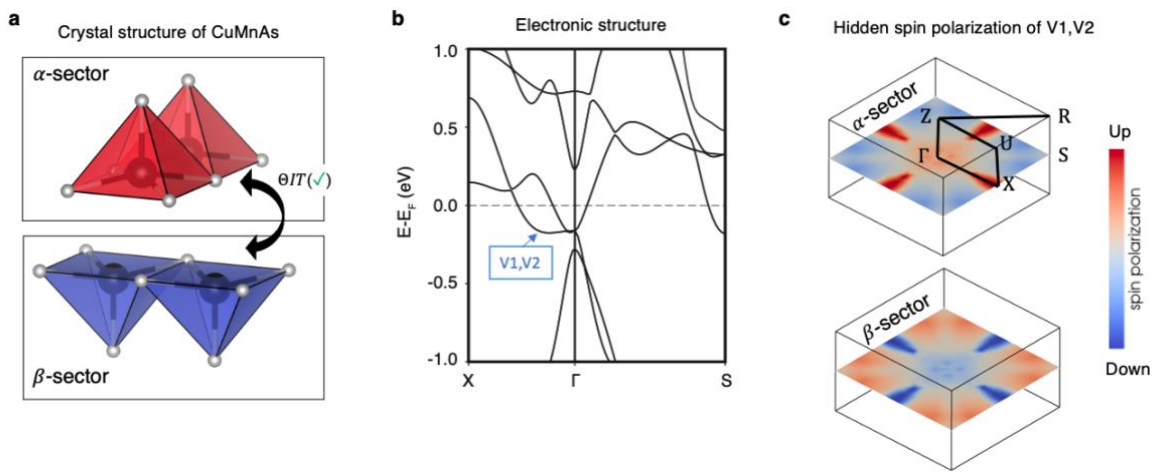

**Figure 3: Hidden spin polarization from individual ferromagnetic sectors in bulk tetragonal CuMnAs (bulk belonging to SST-1 class with sector belonging to SST-5 class). a** Crystal structure of antiferromagnetic CuMnAs composed of two ferromagnetic layers with opposite magnetization (indicated by red and blue polynomials) in the unit cell. The Cu atoms are dismissed. The two layers are referred to as sector-$\alpha$ and sector-$\beta$, respectively; **b** Spin degenerate band structure of CuMnAs; **c** Hidden spin polarization from each individual sector of the highest two valence bands (V1 and V2) on ΓXS k-plane. The up and down spins are mapped to the color from blue to red.

The next step is to validate the predicted hidden spin polarization effect in some of these identified real materials. We studied the sector-projected spin textures on certain wavevector planes for three actual antiferromagnetic materials, CuMnAs[40], $Ca_2MnO_4$[33] and $FeBr_2$[42]. The results are presented below. Additional examples with DFT results are presented in Supplementary Information Section B. These examples proof the existence of the hidden spin polarization effect.

*Hidden spin polarization from individual ferromagnetic sectors*: Figure 3 illustrates the hidden spin polarization effect in tetragonal CuMnAs [40] (bulk belonging to SST-1 class with sectors belonging to SST-5 class). The crystal is antiferromagnetically ordered with its magnetic moments collinearly aligned in the (010) direction. The magnetic space group of the crystal is Pm'mn (MSG type III). The unit cell consists of two MnAs layers ($\alpha$-sector and $\beta$-sector) that are ferromagnetically ordered (Fig. 3a, red and blue color shaped polyhedral are used to indicate oppositely magnetized motifs centered on the magnetic sites). By considering the bulk antiferromagnets as a combination of two alternating non-centrosymmetric sectors ($\alpha$-sector and $\beta$-sector), the material has been demonstrated as a useful platform for electrically switching [24,25] the antiferromagnetic magnetization using the hidden spin polarization from the SOC segregated on each sector. Here, we point out a different SOC-independent scenario that might also be contributing to the observed electric switching in this material, i.e., the Zeeman effect within each ferromagnetic MnAs layer creates a local spin split state anchored on the layer. The two MnAs layers are connected by the $\theta IT$ symmetry which restores the spin degeneracy of the bulk and results in a compensated net spin polarization (Fig. 3b). However, the corresponding spin texture projected onto $\alpha$-sector and $\beta$-sector, shown in Fig. 3c, are persistently aligned in the same direction as its magnetization, therefore, the information is contained in the magnitude of the spin polarization. As shown by the reversed blue and red pattern which are used to map the relative magnitude of the spin up and spin down polarization, the hidden spin polarization is non-zero and is compensated by each other. Examples of hidden spin polarization in spin degenerate bulk antiferromagnets made of spin split ferromagnetic sectors are also illustrated for $CoBr_2$ [42] (bulk belonging to SST-2 with sector belonging to SST-5) and $Ca_3Ru_2O_7$ [45] (bulk belonging to SST-3 with sector belonging to SST-5) in Supplementary Information Section B.

*Hidden spin polarization from individual antiferromagnetic sectors*: Figure 4 illustrates the "hidden spin polarization" effect in antiferromagnetic $Ca_2MnO_4$ [33] (bulk belonging to SST-1 class with sector belonging to SST-4 class). The crystal is antiferromagnetically ordered with its magnetic moments collinearly aligned in the (001) direction. The magnetic space group of the crystal is $I4_1'/a'cd'$ (MSG type III). The unit cell consists of two layers of $MnO_4$ octahedral ($\alpha$-sector and $\beta$-sector) that are antiferromagnetically ordered (Fig. 4a, red and blue color polyhedral are used to indicate oppositely magnetized motifs centered on the magnetic sites). The "magnetic mechanism" [6] within each AFM-ordered sector then creates a local spin split state anchored on the layer. The two $MnO_4$ layers are connected by the $\theta IT$ symmetry which restores the spin degeneracy of the bulk and results in zero net spin polarization (Fig. 4b).

However, the corresponding spin texture projected onto the $\alpha$-sector and $\beta$-sector, shown in Fig. 4c, are persistently aligned in the same direction as its magnetization and are compensated to each other (as indicated by the reversed blue and red pattern which are used to map the relative magnitude of the spin up and spin down polarization). Examples of hidden spin polarization in spin degenerate bulk antiferromagnets made of spin split antiferromagnetic sectors are also illustrated for $MnS_2$ [46] (bulk belonging to SST-2 with sector belonging to SST-4) and $La_2NiO_4$ [47] (bulk belonging to SST-3 with sector belonging to SST-4) in Supplementary Information Section B.

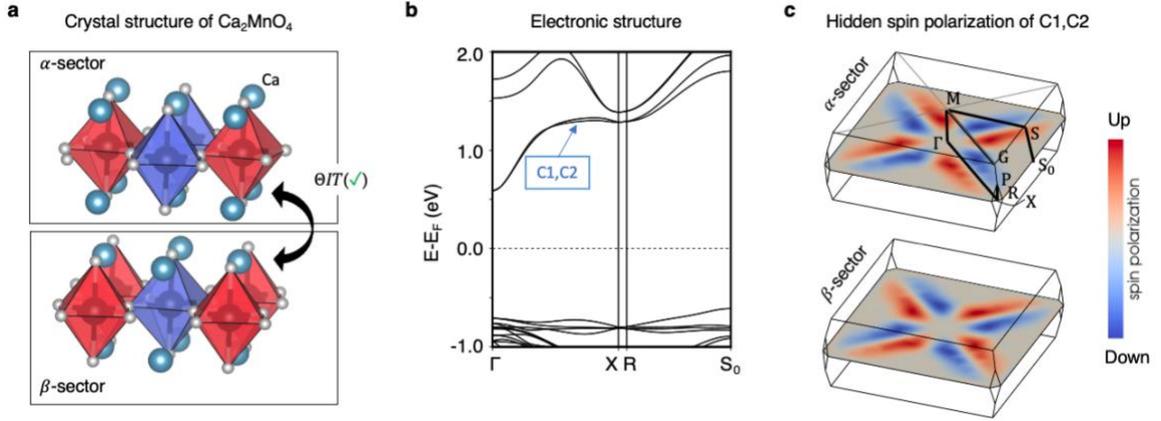

**Figure 4: Hidden spin polarization from the individual antiferromagnetic sector in bulk $Ca_2MnO_4$ (bulk belonging to SST-1 class with sector belonging to SST-4 class).** (a) Crystal structure of antiferromagnetic $Ca_2MnO_4$ is composed of two antiferromagnetic sectors with opposite magnetic ordering (the magnetic ordering is indicated by red and blue polynomials) in the unit cell. The two layers are referred to as sector-$\alpha$ and sector-$\beta$, respectively; (b) Spin degenerate band structure of $Ca_2MnO_4$; (c) Hidden spin polarization from each individual sector of the lowest two conduction bands (C1 and C2) on ΓXR k-plane. The up and down spins are mapped to the color from blue to red.

*Revealing and tailoring the hidden spin polarization by external electric field*: To demonstrate the symmetry connection between local sectors and the subsequent transition from hidden effect to apparent effect mediated by the breaking of the symmetry connection, we apply in our calculations a symmetry-breaking external electric field on an antiferromagnetic compound with hidden spin polarization, hexagonal $FeBr_2$ (DFT settings for applying the electric field is provided in Methods). Figure 5a shows the crystal structure of the bilayer hexagonal $FeBr_2$ (MSG P$_C$-3c1) being a bulk SST-2 class made of sectors belonging to the SST-5 class ($FeBr_2$ ferromagnetic layer). Because the two $FeBr_2$ layers are connected by both $\theta IT$ and $UT$ symmetry, the energy bands of the compounds are exactly spin degenerate. However, the spin degenerate band structure of the SST-2 class $FeBr_2$ (Fig. 5b) is lifted upon the application of an external electric field perpendicular to the layers ($E_z$) – a transition from SST-2 to SST-4. The spin splitting arises because of the external electric field $E_z$ creates a non-equivalent potential on the sectors and breaks the $\theta IT$ and $UT$ symmetry of the bulk that connects the two layers. DFT calculations for different values of the applied field, inserted in Fig. 5b, show that such splitting is linearly proportional to the applied external electric field, but in rates of opposite signs for the bottom conduction bands and the top valence bands. The linear field-dependent splitting suggests the split states are segregated on either layer (sector). Indeed, analysis of the atomic character of the spin split band in Fig. 5c further shows the spin-up (red) bands are dominantly segregated on the $\alpha$-sector, while the spin-down (blue) bands are dominantly segregated on the $\beta$-sector. Therefore, the hidden effect of two-fold degenerate energy states subspace (when $E_z = 0$) can be traced back to the individual $FeBr_2$ layers. We note the hidden spin polarization effect from local "spin-split" sectors has also been recently exemplified and revealed via an electric field in some antiferromagnets [31,48] where external electric field lifts the spin degeneracy. These examples not only verify our understanding of the hidden effect being intrinsic to the bulk but also suggest an external electric field as an effective knob for modulating the hidden effect.

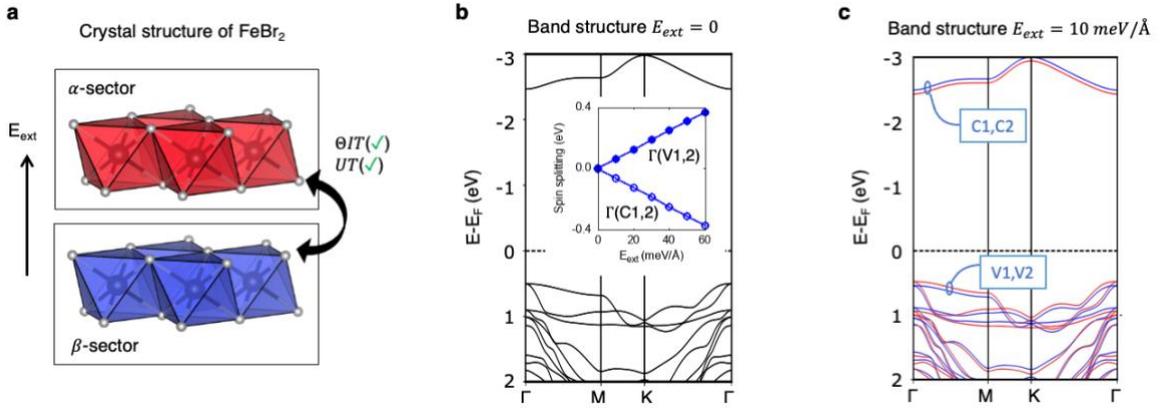

**Figure 5: Revealing the hidden spin polarization in hexagonal $FeBr_2$ using an external electric field.** **a** Crystal and magnetic structure of $FeBr_2$, red and blue shaded polyhedral are used to indicate the oppositely aligned FM $FeBr_2$ layer; **b** spin degenerate band structure of FeBr2 with no external electric field; insert depicts the spin splitting between the top two valence bands and the bottom two conduction bands at Γ as a function of the external electric field; **c** spin split band structure of $FeBr_2$ with a 10 meV/Å z-oriented external electric field. Red and blue lines represent the spin-up and spin-down polarized bands.

## Discussion

*Use magnetic symmetry with SOC to describe the spin splitting of energy bands without SOC:* In collinear antiferromagnetic compounds[26], the existence of *UT* means there is a spatial translation *T* that connects the atomic sites with opposite magnetic moments and keeps the crystal structure invariant. By definition, antiferromagnets with primitive lattice translations that reverse the microscopic magnetic moments are known as having black and white Bravais lattice that is classified as magnetic space group (MSG) type-IV; Antiferromagnets without such translation *T* belongs to MSG type-I and type-III. Such correspondence was also pointed out recently by Ilja Turek [49]. This justifies the use of magnetic space group (with SOC) – therefore avoid the use of the "less familiar" spin symmetry [50] -- to describe the spin splitting of energy bands without SOC. This also allows the use of the tabulated magnetic structure symmetry information provided in material database [32] to sort out candidate materials [30].

*Hidden versus apparent spin polarization in noncollinear antiferromagnets:* While the current paper focuses on the hidden spin polarization in collinear antiferromagnetic compounds, we note that the hidden effect can also exist in noncollinear antiferromagnetic compounds. When a bulk noncollinear antiferromagnetic compound has $\theta IT$ symmetry, the energy bands are spin degenerate. If the system can be further divided into separate sectors that locally violate $\theta IT$, then there could exist hidden spin polarization pertaining to the individual sectors. However, one should note that the symmetry condition of having $UT$ for preserving spin degeneracy in noncollinear antiferromagnetic compounds [30] is not valid anymore, this is because (1) when the spin arrangement is non-coplanar, the MSG type IV does not guarantee the existence of $UT$; Moreover, (2) when the spin arrangement is coplanar, MSG type IV guarantees the existence of $UT$, but the existence of such $UT$ does not always guarantee spin degeneracy. Specifically, when the spin states are not aligned in the same plane of the coplanar plane, the $UT$ symmetry will not reverse the spin states as it works in the collinear magnetic systems. These properties of noncollinear antiferromagnets offer new knobs to tune the hidden versus apparent spin polarization via tilting the local magnetic motifs.

*Experimental detectability:* Analogous to the detection of SOC-induced hidden spin polarization in nonmagnetic compounds (also known as R-2 and D-2 effects) [12], a hidden property can be observed when a probe can resolve the local sectors where the property is not compensated. Specific to hidden spin polarization, the spatial segregation of the spin polarization states allows in principle the detection of the hidden effect in antiferromagnets. Since this effect is intrinsic to the bulk it can be distinguished from the surface effect as the latter sensitively depends on the effective penetration depth of the probing beam [43]. We argue that the recently observed "magnetic splitting" on (001) surface in a collinear antiferromagnetic NdBi where no apparent bulk splitting shall occur attributed by the authors [51] to surface effects are indeed hidden spin polarization from its individual FM sectors that is intrinsic to the bulk. Additional measurement on the (100) surface of NdBi that reveals the hidden spin polarization from the individual AFM sectors where both bulk and surface have no macroscopic magnetic moment might help to clarify the surface *vs* bulk origin of the observed effects. Albeit, for the hidden spin polarization from individual AFM sectors, to detect the AFM spin polarization of the individual sectors, one needs to choose the surface configuration that respects the symmetries of the individual sector that ensure the anti-ferromagnetism of the sector, e.g., mirror plane symmetries perpendicular to the surface plane that connect the spin up and spin down magnetic moments of the AFM sector. Especially, systems with the degenerate states segregated on the different sectors would result in a minimally compensated hidden spin polarization, thus contributing to a robust signal when selectively probing the individual sector, thus being ideal platforms for the detection of the hidden effect.

*Electric field control of the hidden effect*: One of the most desirable features of spin-related phenomena is the possibility of electric and magnetic control. In the case of the hidden spin polarization in AFM, since the unit cell can always be built in terms of two or more sectors, electric fields are a practically direct way of inducing and controlling the existence of spin splitting (as well as its magnitude) via modulating the symmetry relationship between the sectors. For example, in the spin degenerate bulk antiferromagnets made of a pair of spin-split antiferromagnetic sectors (eg. FeSe discussed in Supplementary Information Section C) or ferromagnetic sectors (eg. $FeBr_2$ discussed in the Results Section), external electric field would break the $\theta IT$ and $UT$ symmetry between the spin split sectors, which then implies a transition from hidden effect to apparent effect. Furthermore, the bulk antiferromagnets formed by ferromagnetic layers with alternatively aligned magnetic moments along the direction perpendicular to the ferromagnetic layers (thus hosting hidden spin polarization) could have very different magnetoresistance from the bulk ferromagnets formed by the same ferromagnetic layers but with uniformly aligned magnetic moments. Therefore, switching between the AFM and FM states by external magnetic field could lead to significant change of magnetoresistance, mimicking the tunneling magnetoresistance effect [52]. These perspectives offer electric and/or magnetic means to control the spin-related properties in antiferromagnets.

*Potential application:* Analogous to the application of SOC-induced hidden spin polarization for antiferromagnetic switching in CuMnAs and $Mn_2Au$ [24,25], the SOC-independent hidden spin polarization effects proposed here might also facilitate electrical switching of the antiferromagnetic ordering. Running an electric current through the hidden spin polarized AFM material could induce "hidden" spin transfer torques that alternate in sign on individual sectors of opposite magnetic order originating from the locally spin polarized energy bands. Such "hidden" spin transfer torques may simultaneously switch the magnetic ordering of the individual sectors. The physics might be of particular interest when sectors are well separated quasi-2D ferromagnetic layers and the electric current perpendicularly passes through the layered sectors. Because the perpendicular electric current passes through the layered sectors sequentially, the generated spin current aligned to the magnetization of one layer will tend to reverse the magnetic ordering of the next layer possessing the opposite magnetic ordering when passing through. This offers a possible means to effectively switch the antiferromagnetic ordering of certain "hidden" type antiferromagnetic materials.

## Methods

Electronic structures are calculated using the density functional theory (DFT) method[53-55] with the General Gradient Approximation (GGA)[56,57] implemented in the Vienna Ab initio simulation package (VASP). Structural and magnetic configurations are taken from the MAGNDATA database[32] derived from experiments. The calculations of SOC-independent spin splitting and spin polarization are done using a non-collinear magnetic setting but without the introduction of spin-orbit coupling (i.e., SOC turned off). We adopt the GGA+U method [58] to account for the on-site Coulomb interactions of localized 3d orbitals involved in the calculations. We follow the approach proposed by Neugebauer and Scheffler [59] to apply a uniform electric field to the bilayer slab in the calculations. This approach treats the artificial periodicity of the slab by adding a planar dipole sheet in the middle of the vacuum region.

## Acknowledgement

The work at Colorado University Boulder was supported by the National Science Foundation (NSF) DMR-CMMT Grant No. DMR-2113922. The electronic structure calculations of this work were supported by the US Department of Energy, Office of Science, Basic Energy Sciences, Materials Sciences and Engineering Division under Grant No. DE-SC0010467. This work used resources of the National Energy Research Scientific Computing Center, which is supported by the Office of Science of the US Department of Energy, and the Extreme Science and Engineering Discovery Environment (XSEDE) supercomputer resources, which is supported by National Science Foundation Grant No. ACI-1548562.